\documentclass[aps,pra,reprint,twocolumn,showpacs,superscriptaddress]{revtex4-1}

\usepackage{graphicx}
\usepackage[utf8]{inputenc}
\usepackage{caption}
\usepackage{siunitx}
\usepackage{amsmath}
\usepackage{gensymb}
\usepackage{amssymb}
\usepackage{enumerate}
\usepackage{color}
\usepackage{hyperref}
\usepackage[american]{babel}

\newcommand{\macor}[0]{$\text{MACOR}$}
\newcommand{\off} [0]{{of\!f}}
\newcommand{\on}  [0]{{on}}
\newcommand{\son} [0]{{\,on}}
\newcommand{\Area}[0]{\mathcal{A}}
\newcommand{\orb}[3]{$#1^{#2}\text{#3}$}

\newcommand{\ttP}[0]{\orb{3}{3}{P}}

\newcommand{\ooS}[0]{\orb{1}{1}{S}}

\newcommand{\otS}[0]{\orb{1}{3}{S}}

\newcommand{\tS}[0]{\orb{2}{3}{S}}
\newcommand{\tP}[0]{\orb{2}{3}{P}}

\newcommand{\result}[6]{\parbox{4cm}{\begin{eqnarray*}
	\Area_\on &=& #1 \pm #2 \\ ~~ \Area_\off &=& #3 \pm #4 \\  S &=& #5\% \pm #6\%
\end{eqnarray*}}}

\begin{document}

\preprint{APS/123-QED}

\title{Producing long-lived $2^3\text{S}$ Ps via $3^3\text{P}$ laser excitation in magnetic and electric fields}

\newcommand{\corresponding}[1]{\altaffiliation{Corresponding author, #1}}

\newcommand{\affpolimi}[0]{\affiliation{LNESS, Department of Physics, Politecnico di Milano, via Anzani 42, 22100~Como, Italy}}
\newcommand{\affinfnmi}[0]{\affiliation{INFN, Sezione di Milano, via Celoria 16, 20133~Milano, Italy}}
\newcommand{\affvienna}[0]{\affiliation{Stefan Meyer Institute for Subatomic Physics, Austrian Academy of Sciences, Boltzmanngasse 3, 1090~Vienna, Austria}}
\newcommand{\affinsubria}[0]{\affiliation{Department of Science and High Technology, University of Insubria, Via Valleggio 11, 22100~Como, Italy}}
\newcommand{\affjinr}[0]{\affiliation{Joint Institute for Nuclear Research, Dubna~141980, Russia}}
\newcommand{\affbs}[0]{\affiliation{Department of Mechanical and Industrial Engineering, University of Brescia, via Branze 38, 25123~Brescia, Italy}}
\newcommand{\affinfnpv}[0]{\affiliation{INFN Pavia, via Bassi 6, 27100~Pavia, Italy}}
\newcommand{\afftn}[0]{\affiliation{Department of Physics, University of Trento, via Sommarive 14, 38123~Povo, Trento, Italy}}
\newcommand{\affinfntn}[0]{\affiliation{TIFPA/INFN Trento, via Sommarive 14, 38123~Povo, Trento, Italy}}
\newcommand{\affge}[0]{\affiliation{Department of Physics, University of Genova, via Dodecaneso 33, 16146~Genova, Italy}}
\newcommand{\affinfnge}[0]{\affiliation{INFN Genova, via Dodecaneso 33, 16146~Genova, Italy}}
\newcommand{\affmi}[0]{\affiliation{Department of Physics ``Aldo Pontremoli'', Universit\`{a} degli Studi di Milano, via Celoria 16, 20133~Milano, Italy}}
\newcommand{\affmpi}[0]{\affiliation{Max Planck Institute for Nuclear Physics, Saupfercheckweg 1, 69117~Heidelberg, Germany}}
\newcommand{\afflac}[0]{\affiliation{Laboratoire Aim\'e Cotton, Universit\'e Paris-Sud, ENS Paris Saclay, CNRS, Universit\'e Paris-Saclay, 91405~Orsay Cedex, France}}
\newcommand{\affpolimiII}[0]{\affiliation{Department of Aerospace Science and Technology, Politecnico di Milano, via La Masa 34, 20156~Milano, Italy}}
\newcommand{\affheidelberg}[0]{\affiliation{Kirchhoff-Institute for Physics, Heidelberg University, Im Neuenheimer Feld 227, 69120~Heidelberg, Germany}}
\newcommand{\affcern}[0]{\affiliation{Physics Department, CERN, 1211~Geneva~23, Switzerland}}
\newcommand{\affoslo}[0]{\affiliation{Department of Physics, University of Oslo, Sem Saelandsvei 24, 0371~Oslo, Norway}}
\newcommand{\afflyon}[0]{\affiliation{Institute of Nuclear Physics, CNRS/IN2p3, University of Lyon 1, 69622~Villeurbanne, France}}
\newcommand{\affmoscow}[0]{\affiliation{Institute for Nuclear Research of the Russian Academy of Science, Moscow~117312, Russia}}
\newcommand{\affinfnpd}[0]{\affiliation{INFN Padova, via Marzolo 8, 35131~Padova, Italy}}
\newcommand{\affprague}[0]{\affiliation{Czech Technical University, Prague, Brehov\'a 7, 11519~Prague~1, Czech Republic}}
\newcommand{\affbo}[0]{\affiliation{University of Bologna, Viale Berti Pichat 6/2, 40126~Bologna, Italy}}
\newcommand{\affpv}[0]{\affiliation{Department of Physics, University of Pavia, via Bassi 6, 27100~Pavia, Italy}}
\newcommand{\affnorway}[0]{\affiliation{The Research Council of Norway, P.O. Box 564, 1327~Lysaker, Norway}}
\newcommand{\affheidelbergII}[0]{\affiliation{Department of Physics, Heidelberg University, Im Neuenheimer Feld 226, 69120~Heidelberg, Germany}}
\newcommand{\affbsII}[0]{\affiliation{Department of Civil, Environmental, Architectural Engineering and Mathematics, University of Brescia, via Branze 43, 25123~Brescia, Italy}}

\author{S.~Aghion}
\affpolimi
\affinfnmi

\author{C.~Amsler}
\affvienna

\author{M.~Antonello}
\affinfnmi
\affinsubria

\author{A.~Belov}
\affmoscow

\author{G.~Bonomi}
\affbs
\affinfnpv

\author{R.~S.~Brusa}
\afftn
\affinfntn

\author{M.~Caccia}
\affinfnmi
\affinsubria

\author{A.~Camper}
\affcern

\author{R.~Caravita}
\corresponding{ruggero.caravita@cern.ch}
\affcern

\author{F.~Castelli}
\affinfnmi
\affmi

\author{G.~Cerchiari}
\affmpi

%

\author{D.~Comparat}
\afflac

\author{G.~Consolati}
\affpolimiII
\affinfnmi

\author{A.~Demetrio}
\affheidelberg

\author{L.~Di~Noto}
\affge
\affinfnge

\author{M.~Doser}
\affcern

\author{C.~Evans}
\affpolimi
\affinfnmi

\author{M.~Fan\`{i}}
\affge
\affinfnge
\affcern

\author{R.~Ferragut}
\affpolimi
\affinfnmi

\author{J.~Fesel}
\affcern

\author{A.~Fontana}
\affinfnpv

\author{S.~Gerber}
\affcern

\author{M.~Giammarchi}
\affinfnmi

\author{A.~Gligorova}
\affvienna

\author{F.~Guatieri}
\corresponding{francesco.guatieri@unitn.it}
\afftn
\affinfntn

\author{P.~Hackstock}
\affvienna

\author{S.~Haider}
\affcern

\author{A.~Hinterberger}
\affcern

\author{H.~Holmestad}
\affoslo

\author{A.~Kellerbauer}
\affmpi

\author{O.~Khalidova}
\affcern

\author{D.~Krasnick\'y}
\affge
\affinfnge

\author{V.~Lagomarsino}
\affge
\affinfnge

\author{P.~Lansonneur}
\afflyon

\author{P.~Lebrun}
\afflyon

\author{C.~Malbrunot}
\affcern
\affvienna

\author{S.~Mariazzi}
\afftn
\affinfntn

\author{J.~Marton}
\affvienna

\author{V.~Matveev}
\affmoscow
\affjinr

\author{Z.~Mazzotta}
\affinfnmi
\affmi

\author{S.~R.~M\"{u}ller}
\affheidelberg

\author{G.~Nebbia}
\affinfnpd

\author{P.~Nedelec}
\afflyon

\author{M.~Oberthaler}
\affheidelberg

\author{D.~Pagano}
\affbs
\affinfnpv

\author{L.~Penasa}
\afftn
\affinfntn

\author{V.~Petracek}
\affprague

\author{F.~Prelz}
\affinfnmi

\author{M.~Prevedelli}
\affbo

\author{B.~Rienaecker}
\affcern

\author{J.~Robert}
\afflac

\author{O.~M.~R{\o}hne}
\affoslo

\author{A.~Rotondi}
\affinfnpv
\affpv

\author{H.~Sandaker}
\affoslo

\author{R.~Santoro}
\affinfnmi
\affinsubria

\author{L.~Smestad}
\affcern
\affnorway

\author{F.~Sorrentino}
\affge
\affinfnge

\author{G.~Testera}
\affinfnge

\author{I.~C.~Tietje}
\affcern

\author{M.~Vujanovic}
\affcern

\author{E.~Widmann}
\affvienna

\author{P.~Yzombard}
\affmpi

\author{C.~Zimmer}
\affcern
\affmpi
\affheidelbergII

\author{J.~Zmeskal}
\affvienna

\author{N.~Zurlo}
\affinfnpv
\affbsII

\collaboration{The AEgIS collaboration}
\noaffiliation{}


\date{\today}


\begin{abstract}
Producing positronium (Ps) in the metastable \tS{} state is of interest for various applications in fundamental physics. We report
here about an experiment in which Ps atoms are produced in this
long-lived state by spontaneous radiative decay of Ps excited to
the \ttP{} level manifold. The Ps cloud excitation is obtained with a UV laser
pulse in an experimental vacuum chamber in presence of guiding
magnetic field of $\SI{25}{\milli\tesla}$ and an average electric field of $\SI{300}{\volt\per\centi\meter}$.
The indication of the \tS{} state production is obtained from a novel
analysis technique of single-shot positronium annihilation lifetime
spectra. Its production efficiency relative to the total amount of formed Ps is evaluated by fitting a simple rate equations model to the experimental data and found to be $ (2.1 \pm 1.3) \, \% $.
\end{abstract}

\maketitle{}

\section{Introduction} \label{SectionI}

Positronium (Ps) is the hydrogen-like bound state of an electron and its antiparticle, the positron. Ps can exist in two different ground states: the singlet state p-Ps, rapidly decaying into two photons with a lifetime of \SI{0.125}{\nano\second}, and the triplet state o-Ps, annihilating in 3$\gamma$ photons with a lifetime of \SI{142}{\nano\second} in vacuum \cite{PositroniumReview}. Being a purely leptonic two-body system, this short-lived atom constitutes a privileged opportunity for high-precision studies of Quantum Electrodynamics (QED) for bound states, to test the validity of accurate QED correction calculations and to measure with high accuracy the fine structure constant $ \alpha $, the Rydberg constant $ R_\infty $ and the electron/positron mass $ m_e $ \cite{Karshenboim}. As a matter of fact, the Ps energy levels can be calculated as a perturbative series in powers of $ \alpha $ with very high precision, only limited by the knowledge of the fundamental constants \cite{QEDexpansion}. Hence, Ps offers a great advantage with respect to analogous calculations and experiments on hydrogen, for example, because in this case QED predictions require a precise knowledge of the proton finite-size effects \cite{Karshenboim,Karshenboim2} which are increasingly important in H spectroscopy.

Some experimental tests are based on precision laser spectroscopy \cite{Karshenboim2}, in particular exploiting the two-photon Doppler-free transition \otS{} -- \tS{} between long-lived triplet o-Ps energy levels. These experiments can be realized because of the metastability of the \tS{} state, which decays, in a field-free region, by annihilation in \SI{1.14}{\micro\second} (the lifetime is increased by a factor of eight compared to the ground state due to the decrease in the overlap of the positron--electron wave-function \cite{Charlton}) with the consequence of a very narrow natural radiative width. First observations of this optical excitation were performed by Chu \emph{et al.} \cite{Chu} by using pulsed lasers on Ps atoms exiting a metallic surface. In improved experiments using continuous-wave lasers, the transition frequency was measured with a precision of 2.6 ppb, sufficient to provide a test for $ o(\alpha^4) $ QED corrections \cite{Fee}. To verify calculations to higher order, further experiments based on advanced Ps synthesis starting from a continuous positron beam transported towards a silica converter \cite{Alberola}, and using an enhanced laser excitation system, has been developed: new accurate determinations of the \otS{} -- \tS{} transition frequency are expected from a combination of Ps detection techniques, also using the observation of the \tS{} state annihilation \cite{Crivelli}. Other QED tests are based upon the determination of the Ps $ n = 2 $ fine structure splitting, in particular using microwaves to induce electric-dipole transitions between the metastable \tS{} state (produced by impinging positron beams on metal surfaces \cite{metalsurf}) and the \tP{} sublevels \cite{Hatamian}. All these Ps experiments can benefit from an efficient and clean production of the \tS{} metastable state, alternative to the usual methods of Ps excitation via a two-photon process or by direct excited state emission from Ps converters.

Another important research field in which Ps plays a privileged role is the study of matter-antimatter gravitational interactions. A direct measurement of the gravitational free fall of antimatter atoms, or matter-antimatter neutral systems like Ps, is foreseen as a test for speculative models aiming to describe the observed asymmetry between matter and antimatter in the Universe by a gravitational asymmetry \cite{Nieto}. Specifically, for pursuing this objective, measurements made on Ps atoms are complementary to other experiments actually proposed \cite{Gbar, AlphaG} or running \cite{Kellerbauer, Aegis} employing antihydrogen; in the former case one expects that the possible absence of a free-fall can be a clear signature of an ``antigravity'' component in the gravitational interaction violating the weak equivalence principle. Experimental proposals currently under evaluation are based on detecting the free fall of Ps atoms, either by guiding a Rydberg Ps beam towards a position-sensitive detector with electrostatic potentials \cite{MillsCassidy} or by accurately measuring the vertical displacement of the atoms' trajectory with a matter-wave optical interferometer \cite{Oberthaler} or with a matter-wave mechanical interferometer in a novel Talbot-Lau configuration for increased compactness \cite{Sala}.

Whichever experimental layout will be found most suitable, it will anyway be necessary to prepare a sample of Ps atoms in a long-lived state, otherwise the rapid annihilation would quickly reduce their useful number making it very difficult, for instance, to distinguish an interferometric pattern against the background. 

Laser excitation to long-lived Rydberg levels has been the first studied solution to overcoming the lifetime limitation \cite{MillsCassidy, Aegis}. This promising route, however, has some potential limitations. It requires an experimental apparatus essentially free from residual fields (and field gradients) in the free-fall region, because of the high electrical polarizability of Rydberg states and their significant ionization rate due to the motional Stark effect induced by magnetic fields. Moreover, laser excitation and subsequent spontaneous optical decay populate a large number of sublevels, making the control of such detrimental effects difficult. 

An alternative way to produce long-lived samples of Ps atoms is indeed laser excitation of the \tS{} metastable level. A collimated beam of metastable Ps atoms has been shown to be extremely useful for improving inertial sensitivity in proposed matter-wave interferometric layouts \cite{Oberthaler, Sala}. 


Recently, the production of Ps atoms in the \tS{} state by single-photon excitation in the presence of a static electric field was demonstrated \cite{AlonsoHoganCassidy}. This result was achieved because of the Stark mixing between S and P sublevels induced by an electric field of a few \si{\kilo\volt\per\centi\meter}, which allows to momentarily populate also the \tS{} states by a laser pulse resonant with the (electric-dipole-allowed) \otS{} -- \tP{} transition. Subsequently, to avoid the rapid radiative decay of the mixed states towards the ground state, the electric field was adiabatically reduced to zero, finally leaving a beam of Ps metastable atoms. In this experiment a $ 6.2 $ \% production efficiency relative to the number of formed Ps atoms was reported, consistent with the expected losses due to the electric field switching time.

Another possible route for the production of Ps metastable state, which is conceptually simpler and ideally free from this drawback, is based on the fact that the desired \tS{} states can be conveniently populated by spontaneous radiative decay from o-Ps atoms laser-excited from the ground staet to the $ n = 3 $ level (precisely on \ttP{} states). This decay competes with the radiative decay to the triplet ground staet \otS{}, and in fact the expected value for the branching ratio of the spontaneous radiative decay \ttP{} -- \tS{} (in absence of magnetic and electric fields) amounts to 12\% \cite{Villa,Caravita}. For a comparison, the theoretical maximum efficiency of \tS{} production by the two-photon Doppler free excitation process was determined to be $ 17.6\% $ \cite{Haas}. The pathway above proposed was experimentally demonstrated in the hydrogen system \cite{Harvey}, where a beam of $ 10^6 $ metastable atoms/s was obtained. Note that scaling considerations between the hydrogen atom and Ps lead to the conclusion that their branching factors must be identical.

Laser excitation of the \ttP{} Ps levels manifold by UV pulses was recently achieved \cite{NEqual3} in the framework of the AEgIS experimental program devoted to antihydrogen synthesis for gravitational studies \cite{Aegis}. In this work we used the same apparatus to demonstrate the feasibility of producing a source of metastable or long-lived Ps atoms. The present experiment was performed in a dedicated chamber where both guiding magnetic and electric fields were present. The detection technique is based on the analysis of the single-shot positron annihilation lifetime spectroscopy (SSPALS) spectrum \cite{Spectroscopy,TrapBasedBeam}.

\begin{figure}[h!tp]
\centering
\includegraphics[width=0.8 \linewidth]{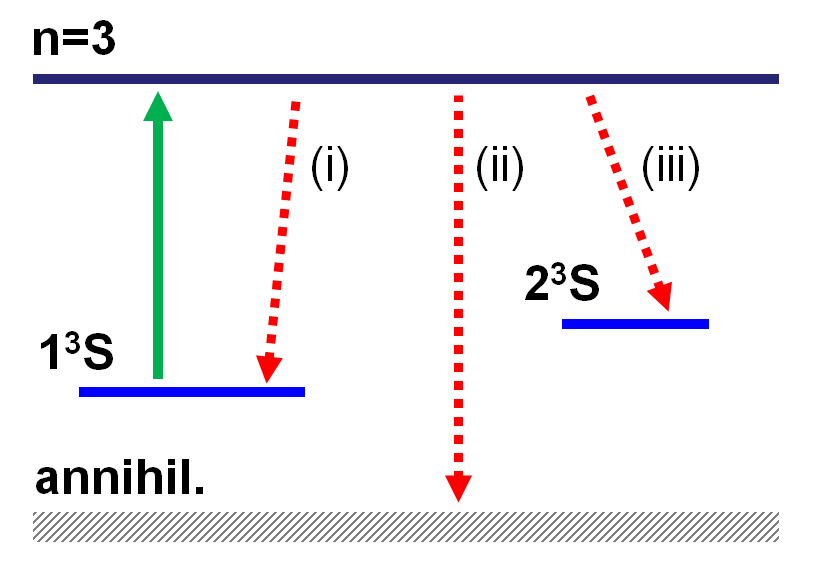}
\caption{Partial energy level diagram of Ps with the $ n = 3 $ laser excitation (solid arrow) and subsequent de-excitation processes (dotted arrows) at play in our experiment (see text).}
\label{simpl_scheme}
\end{figure}

A simple illustration of the excitation and de-excitation processes involved in our experiment is shown in Fig.~\ref{simpl_scheme} (see Section IV for a more detailed discussion). Ps atoms produced in a triplet o-Ps ground state are excited by a nanosecond UV laser pulse to the $n = 3$ manifold, whose sublevels are mixed by the presence of the guiding electric field. The excited atoms then follow three main de-excitation paths:
(i) a first fraction of atoms decays spontaneously back to the \otS{} triplet ground state in a few tens of nanoseconds, where they start again the normal ground state annihilation path;
(ii) a second fraction of atoms quenches in the $\SI{25}{\milli\tesla}$ magnetic field present in the experimental chamber and decays to the \ooS{} singlet ground state where they promptly annihilate;
(iii) a third fraction decays to the metastable \tS{} state, as discussed above. 

In the case of the first decay channel no significant change in the overall lifetime of the Ps sample is introduced by the laser excitation. Consequently, the shape of the corresponding annihilation signal in SSPALS spectra is left almost unvaried by the presence of the laser. The second decay channel causes only an immediate loss of a small fraction of Ps atoms occuring simultaneously with the laser shot, and a consequent relative reduction of the population of Ps atoms in \otS{} at later times. This causes a small signal reduction in SSPALS spectra (of the order of $ 2 - 3 \, \% $, see \cite{NEqual3}) but does not alter the signal shape by introducing delayed annihilation signals. The third decay channel is the one of interest here for obtaining long-lived metastable Ps atoms. In the absence of an electric field, this state does not have first order dipole allowed radiative decay paths and it only annihilates with the long annihilation lifetime of $\SI{1.14}{\mu s}$ \cite{Charlton}. Hence this would constitue a long-lived component present in the cloud of Ps excited atoms, and in the SSPALS spectrum one would observe a decrease of the annihilation signal immediately after the laser shot, followed by an increase at very later times when the Ps atoms hit the walls of the experimental chamber.

In the presence of an electric field (as in our experimental setup), Stark mixing considerably reduces the lifetime of Ps in the \tS{} state, as its lifetime is affected by the partial mixing with the \tP{} states which can radiatively decay back to \otS{} (the spontaneous radiative decay lifetime of \tS{} in our field configuration is indeed $ \sim \SI{105}{\nano\second} $, see the detailed discussion in Section \ref{SectionIV}). However, even if its lifetime is reduced by the presence of the fields, Ps in the \tS{} state still constitues a long-lived component, compared to the triplet ground state \otS{}, which alters SSPALS spectra at later times.

A novel analysis technique of SSPALS data has been developed to highlight the appearance of slightest modifications in these
spectra which can be induced by the presence of a small
fraction of long-lived Ps states. This novel technique can be
of general interest for data analysis in similar experiments.
In the following sections, we describe the experimental apparatus, present the statistical technique used for the accurate and reliable analysis of the results, and compare them with a simple rate equation model describing radiative and annihilation decays well after the laser excitation.

\section{Experimental methods} \label{SectionII}

The system used to perform the present experiment is described in detail elsewhere \cite{NEqual3, BunchingSystem, PositronBunching}. Briefly, positrons emitted by $\beta^+$ decay of a $\SI{9}{mCi}$ $\ ^{22}\text{Na}$ source were slowed by a solid Ne moderator \cite{SolidNeonModerator} to a kinetic energy of a few \si{\electronvolt}, trapped and cooled in a Surko-style trap through the use of buffer gas \cite{PlasmaAndTraps}. The cooling efficiency of the Ne moderator decreases by aging, peaking when the moderator is grown (\emph{i.e.} it was heated and evaporated and then regenerated by condensation of gaseous Ne) then declining with an initial rate of several \% of efficiency per hour lost during the first few hours \cite{NeModerator}.

Subsequently positrons were moved to a second trap (accumulator) where several pulses from the first trap were stored. There the positron plasma was radially compressed using the rotating-wall technique \cite{PositronCompression} and then extracted by fast pulses of the electric potential on the trap electrodes in the form of $ \SI{20}{\nano\second} $ bunches of $ \sim 3 \cdot 10^7 $ positrons at \SI{100}{\electronvolt} axial energy. The cloud was then transported to a magnetic-field-free region where it was further compressed in time \cite{PositronTimeBunching} using a 24-electrode buncher \cite{PositronBunching} to about $ \SI{7}{\nano\second} $ and accelerated onto a nanochanneled silicon target with a final kinetic energy of $\SI{3.3}{keV}$.

In the target, $ e^+ $ were efficiently converted into o-Ps and emitted into vacuum \cite{NCP, CryoNCP}. A calibrated CsI detector and a microchannel plate (MCP) with a phosphor screen, set in place of the target, were used to characterize the number and the spot dimension of positrons impinging on the target. It was estimated that $ 30-40\% $ of positrons released from the accumulator hit the target in a spot with a full width at tenth maximum $\leq\SI{4}{\milli\meter}$, the rest being lost in the transfer. Two symmetric coils generating a magnetic field perpendicular to the target were used to increase the positron transport efficiency onto the target. The described experiments were performed while keeping the target at room temperature in a \SI{25}{\milli\tesla} magnetic field environment and in the presence of an electric field of around \SI{300}{\volt\per\centi\meter}, mostly parallel to the magnetic field in the laser excitation region. This field was produced by the last electrode of the buncher that acts as an electrostatic lens.

\begin{figure}[htp]
\centering
\includegraphics[width=\linewidth]{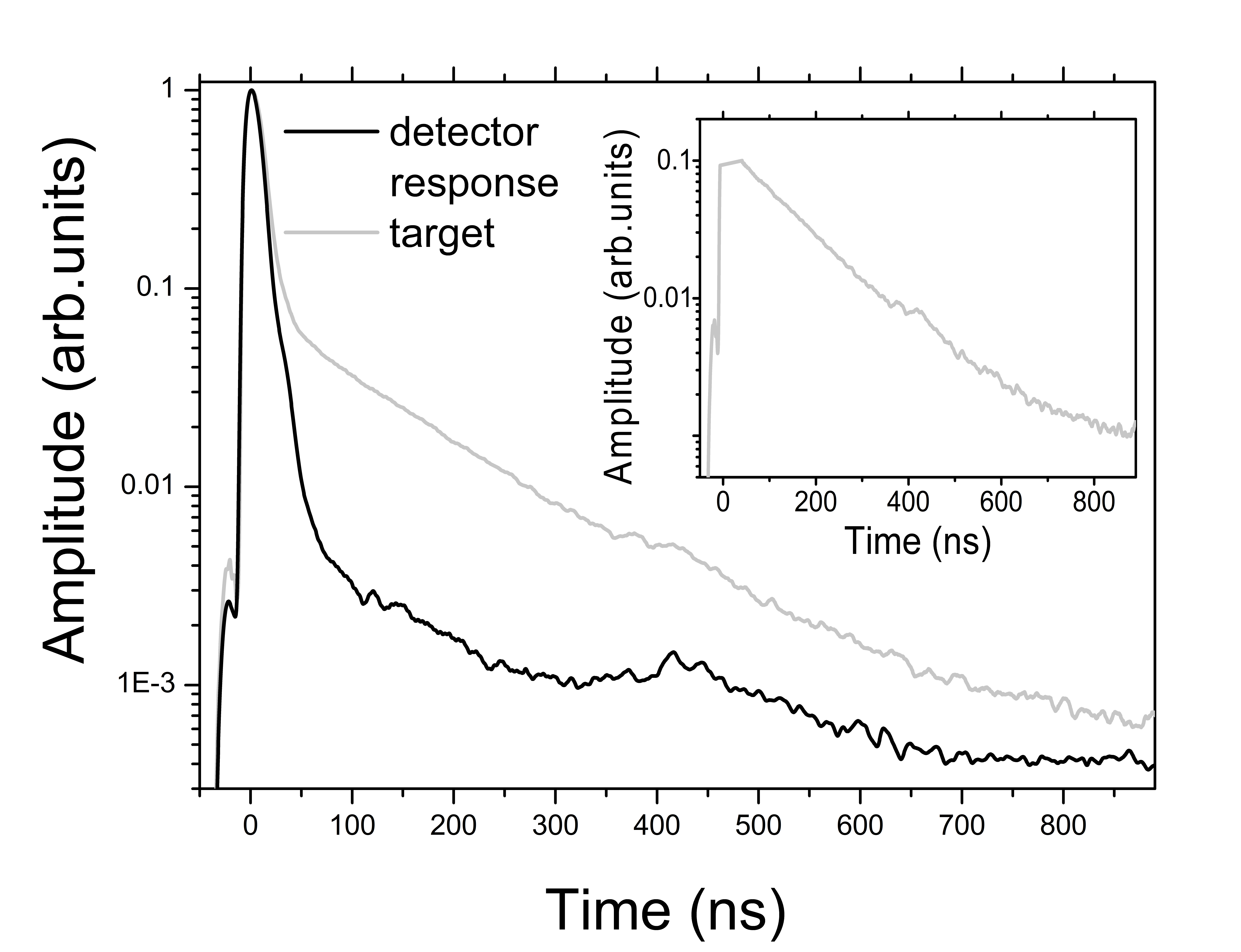}
\caption{SSPALS spectra measured in absence of Ps formation (black line, average of 338 shots), and in presence of Ps formation (gray line, average of 159 single shots) normalized to the peak height. In the inset, the SSPALS acquired on the target with the high gain channel is shown. The time origin is taken at the maximum of the prompt peak.}
\label{Background}
\end{figure}

As mentioned above, Ps annihilations were monitored using the SSPALS technique. A $20 \times 25 \times 25 \SI{}{\milli\meter}$ lead tungstate (PbWO$_4$) scintillator \cite{TrapBasedBeam} coupled to a Hamamatsu R11265-100 photomultiplier tube (PMT) was placed $\SI{40}{\milli\meter}$ above the target to record photons emitted by positron-electron annihilations. To enhance the resolution at the longest decay times, the signal from the PMT was split and sent to two channels of a $\SI{2.5}{GHz}$ oscilloscope with high ($\SI{100}{mV/division}$) and low ($\SI{1}{V/division}$) gain. Joined data from the two channels give the SSPALS spectrum shown in Fig.~\ref{Background} when $e^+$ are bunched on the surface of the MCP (no Ps formation; detector response) and on the target (Ps formation). In the presence of Ps formation SSPALS spectra present a prompt peak and a tail. The prompt peak, in a region up to $\SI{100}{\nano\second}$ from the positron implantation, is due to $2\gamma$ annihilations of $e^+$. The tail is dominated by the Ps decay in vacuum, therefore it is proportional to the time derivative of the $n=1$ Ps population $-dN_1/dt$. In the inset the signal given by the high gain channel on the target is reported. Calibration measurements have shown that the detector performs linearly in the dynamic range of the high gain channel but could show slightly non linear behaviors in the peak region within the dynamic range of the low gain channel.

For the present measurements a first UV laser pulse was used for Ps excitation to the $n = 3$ energy levels. A second IR laser pulse was also employed for selective photoionization of the excited atoms. The laser setup is described in detail in Refs. \cite{TwoStepExcitation} and \cite{LaserApparatus}, and was previously used to demonstrate $n=3$ excitation of Ps \cite{NEqual3}. With respect to that setup, here the laser system was improved to deliver about 2.5 times the energy in the UV. The UV pulse energy was kept above \SI{80}{\mu J}, peaking at about \SI{130}{\mu J} in optimal conditions (measured outside the experimental chamber, $5 \%$ absorption of the viewport not considered). The wavelength of the UV laser was monitored during the measurements to verify that unavoidable thermal drifts induced by prolonged operation of the pump laser did not alter the wavelength setting on the resonance of the $1\text{S} \rightarrow 3\text{P}$ transition ($\lambda = 205.045 \pm 0.005 \si{\nano\meter}$) \cite{NEqual3}. It had a horizontal polarization (\emph{i.e.} polarization perpendicular to the sample), an asymmetric, nearly Gaussian temporal profile with a  full-width at half maximum (FWHM) of \SI{1.5}{\nano\second}, and a Gaussian-like spectral profile with $ \sigma_{UV} = 2 \pi \cdot \SI{48}{\giga\hertz} $. 

\begin{figure}[htp]
	\centering
	\includegraphics[width=0.45 \linewidth]{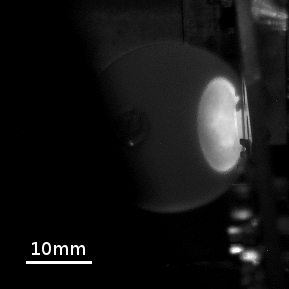}
	\includegraphics[width=0.45 \linewidth]{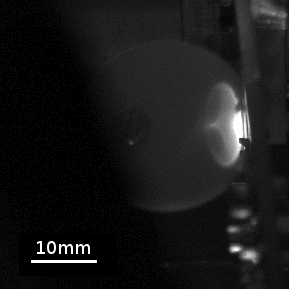}
	\caption{Intensity distribution in front of the target  as observed on the \SI{25.4}{\milli\meter} \macor screen rotated by $45 \degree$ downwards, by a camera on a $45 \degree$ angle viewport (with respect to the beam direction) and orthogonal to the screen. This beam has an elliptical shape and an asymmetric distribution of light due to the passage in an astigmatic telescope. The left panel shows the intensity distribution without the telescope. The right panel shows the intensity distribution with the lens tilted by about $10 \degree$, showing the increased intensity in front of the active area. The exposure settings of the acquisition camera were kept the same.}	
	\label{intensity}
\end{figure}

Here the increase in energy of the UV opened the possibility to sacrifice about $20\%$ by adding a telescopic system to control the beam spot size and shape. The spatial intensity distribution was almost Gaussian before entering the last telescope, where the two lenses were used to significantly reduce the spot size and to make it astigmatic so that most of the energy ends up in front of the active region of the target (see Fig.~\ref{intensity}). The effective size of the UV beam used during the measurement was about 3.0 - 3.5 \si{\milli\meter} FWHM both in horizontal and vertical direction. 

The second, intense infrared (IR) laser pulse at $\SI{1064}{\nano\meter}$ was simultaneously delivered to the experimental chamber to selectively photoionize o-Ps in the $n =3$ excited state. This horizontally polarized pulse had an energy of $\SI{50}{\milli\joule}$ and a temporal FWHM of $\SI{6}{\nano\second}$. It was superimposed on the UV pulse both in time with a precision of $ < \SI{1}{\nano\second}$, by using an optical delay line, and in space by increasing its size so as to completely cover the excitation pulse area (top-hat profile of $\geq \SI{20}{\milli\meter}$ diameter). Both beams were aligned on the target region by monitoring their position with a CCD camera on a \macor screen placed inside the vacuum region, a few cm away from the target, rotated $45 \degree$ downwards so as to face the camera, which was placed on a $45 \degree$ angled viewport. A mutual synchronization of positrons and laser pulses with a time resolution of $\SI{2}{\nano\second}$ and a jitter of less than $\SI{600}{\pico\second}$ was obtained by a custom field-programmable gate array (FPGA) synchronization device (see \cite{LaserApparatus}). The time delay between the prompt positron annihilation peak and the laser pulses was set to $\SI{16}{\nano\second}$.

\section{Measurements and analysis of the results} \label{SectionIII}

Two different sets of measurements were performed. The first one was carried out by simultaneously firing the UV and IR lasers. The goal was to verify that the excitation of Ps to the $n = 3$ level manifold and the following photoionization only induce a proportional decrease of the o-Ps population decaying to three $ \gamma $ rays and no significant increase of the annihilations at later times. The second set of measurements was performed by firing only the UV laser in order to populate the long-lived \tS{} state and observe whether an excess of signal was present at later times, induced by its decay (the annihilation channel (iii) discussed in the Introduction).

\begin{figure}[htp]
	\centering
	\includegraphics[width=\linewidth]{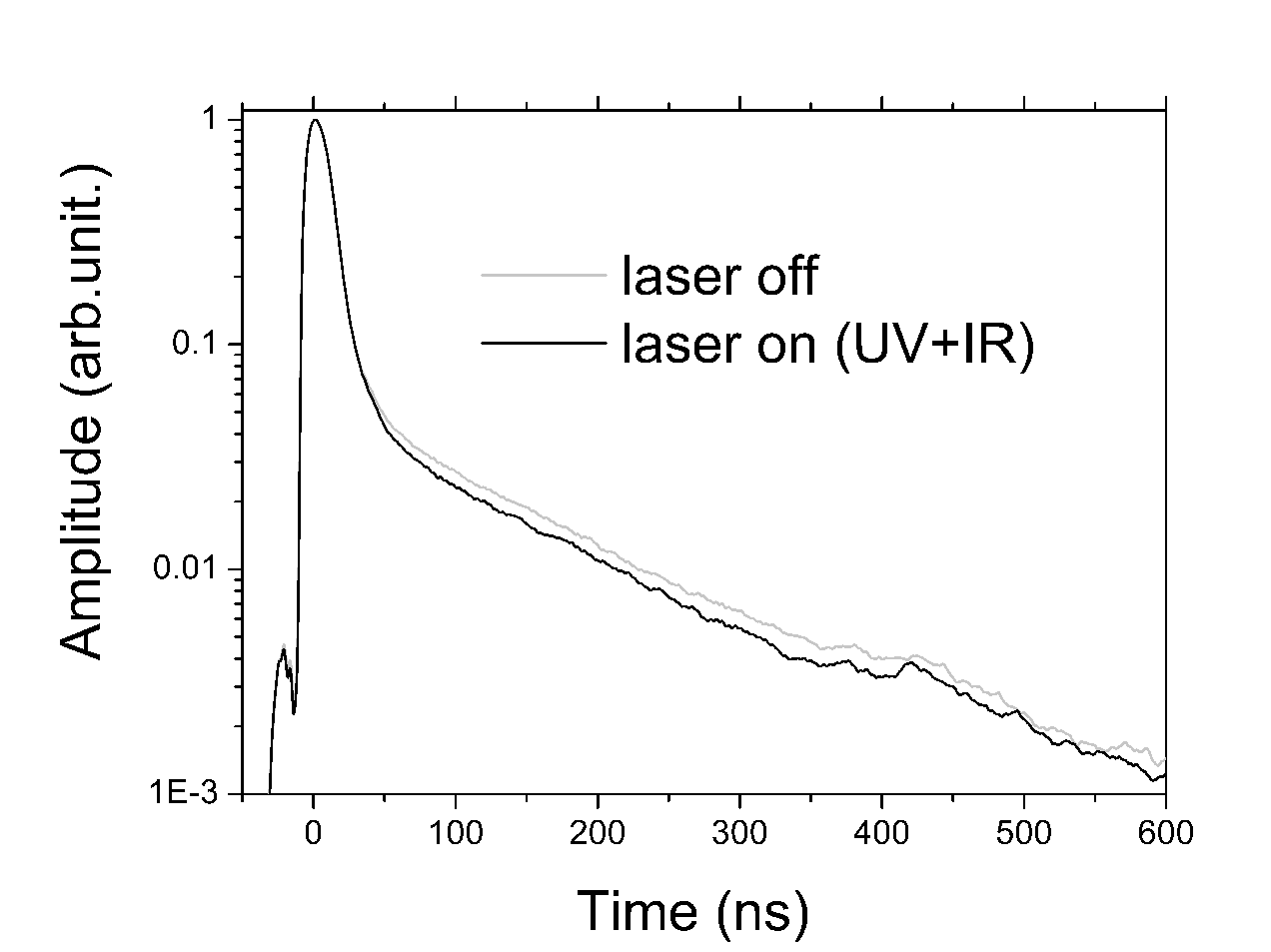}
	\caption{SSPALS spectra of Ps into vacuum with lasers off in gray and UV+IR lasers on ($205.045 + 1064 \SI{}{\nano\meter}$) in black normalized to the peak height. Each spectrum is the average of 159 single shots. The laser pulses were shot $\SI{16}{\nano\second}$ after the prompt positron annihilation peak. The time origin is taken at the maximum of the prompt peak.}
	\label{SSPALS_UVIR}
\end{figure}

The comparison between the averaged SSPALS spectra acquired with lasers off and both UV+IR lasers on is reported in Fig.~\ref{SSPALS_UVIR}. Photoionization of the \ttP{} state dissociates the Ps, and the free positrons are quickly accelerated back toward the last negative electrode of our setup, where they annihilate in a few \si{\nano\second}. Therefore these positrons do not contribute to the delayed annihilations in the SSPALS spectrum.

\begin{figure}[h!tp]
	\centering
	\includegraphics[width=\linewidth]{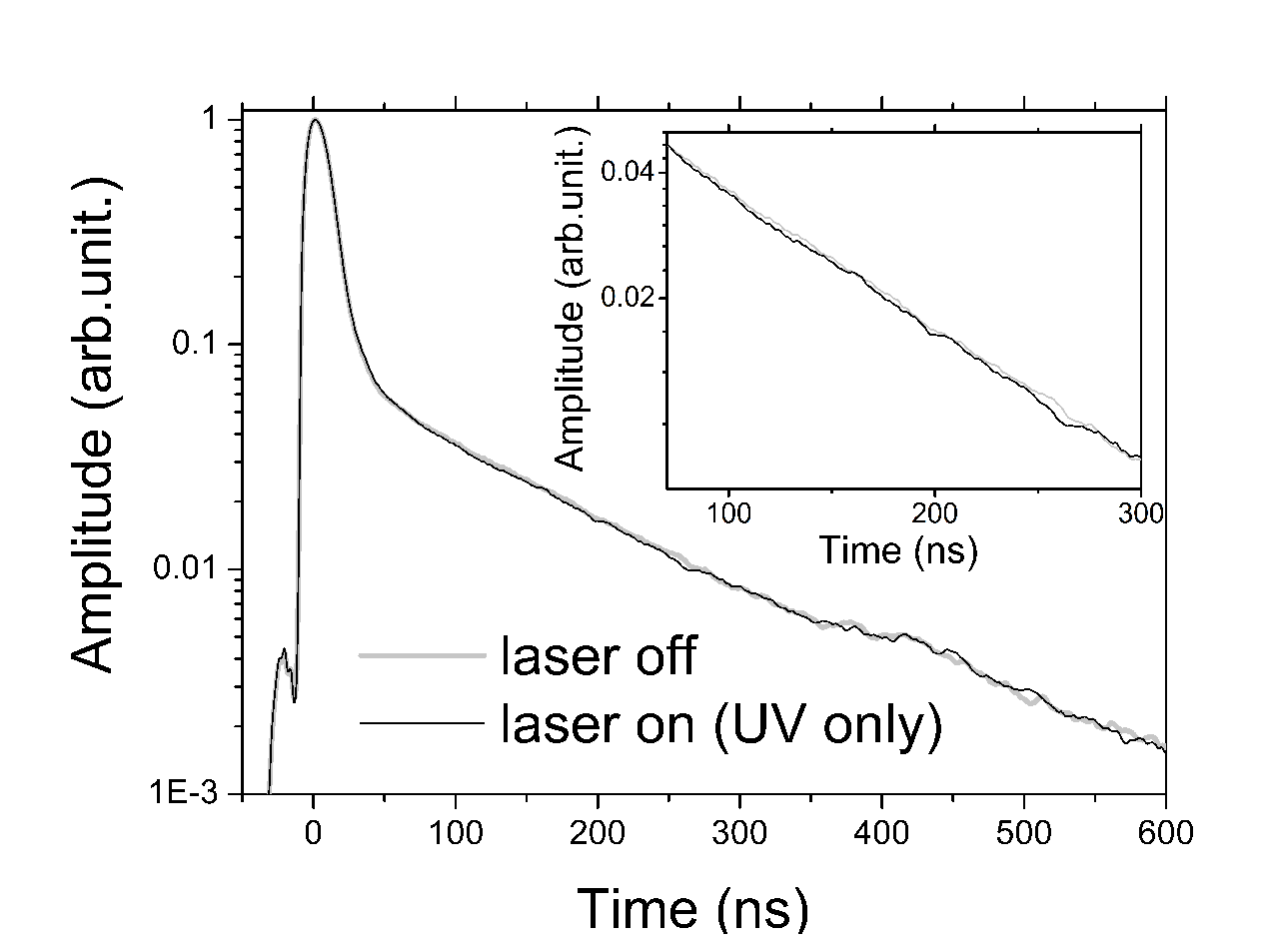}
	\caption{SSPALS spectra of Ps into vacuum with lasers off in gray and UV laser on ($\SI{205.045}{\nano\meter}$) in black normalized to the peak height. Each spectrum is the average of 179 single shots. The laser pulses were shot $\SI{16}{\nano\second}$ after the prompt positron annihilation peak. The detail of the difference of the two SSPALS spectra between $70$ and $\SI{300}{\nano\second}$ from the prompt peak is reported in the inset. The time origin is taken at the maximum of the prompt peak.}
	\label{SSPALS_UV}
\end{figure}

The comparison between the averaged SSPALS spectra acquired with lasers off and only the UV laser is shown in Fig.~\ref{SSPALS_UV}. With reference to the annihilation dynamics described in the Introduction, the quenching of the excited states in the \SI{25}{\milli\tesla} magnetic field (channel (ii)) and, moreover, the decay to the metastable \tS{} state (channel (iii)), are expected to induce a decrease of the Ps annihilations immediately after the UV shot, and an increase at later time with respect to the SSPALS spectra with both lasers off (see also inset in Fig.~\ref{SSPALS_UV}).

The measurement was carried out by regenerating the Ne moderator and then alternating shots with the UV laser on to shots with the laser off up until the moderator aging caused the positron beam intensity to decrease enough to justify the growth of another moderator (usually when the signal reached about 75 \% of the initial value). Positron accumulation time limits the shot repetition rate to about three minutes; each sequence of shots taken between two successive moderator regenerations is typically composed of between 16 and 25 shots. 
A total of $ 179 $ shots with the UV laser on, $ 159 $ shots with both the UV and IR lasers on and $ 338 $ with both lasers off were acquired. 

A way to quantify the amount of delayed annihilations induced by the presence of the laser is to compute the relative difference of the areas of SSPALS spectra with/without laser in selected time windows. We adopt here the same analysis methodology already demonstrated for detecting long-lived Rydberg states of Ps \cite{NEqual3} based on the calculation of $ S $ parameter, which is defined as $ S_i = (\Area_i^\off - \Area_i^\on) / \Area_i^\off $ where $ \Area_i^\on $ is the area under the spectrum in a selected time window of a single shot when the laser is on and $ \Area_i^\off $ is the area of the following shot in the same time interval when the laser is blocked.
This definition of $S_i$ implicitly assumes the same positron beam intensity for both shots, or a proper normalization of the spectra, because the fraction of emitted Ps is expected to be proportional to the implanted positron number.

Any quantity proportional to the positron beam intensity can provide a valid normalization tool. The prompt peak height is a frequently adopted choice \cite{NEqual3, PeakH}. For the present case, however, we cannot rely on it as we aim to higher precision in normalizing the spectra than the level of linearity of the scintillator detectors in the prompt peak (as mentioned before).
To circumvent this difficulty, we used only the portion of the SSPALS spectrum acquired at high gain (therefore avoiding any detector saturation, see inset in Fig.~\ref{Background}) to compute a suitable normalization factor. This required the development of a novel normalizing technique of SSPALS spectra. We will name it \emph{detrending} for its conceptual similarity with the homonymous technique used in signal analysis \cite{DetrendingDNA}. The technique here developed was specifically optimized for the analysis of SSPALS measurements where two distinct class of time-interleaved measurements are present, in order to make the most efficient use of the a-priori knowledge about the laser status to reduce the error in estimating the normalization factor. A detailed and general formulation of our technique can be found in the Appendix.

The technique consists firstly in computing the area under the SSPALS spectrum in a suitable time window, then pairing it with the time elapsed from the last moderator regeneration to the time the specific shot had been acquired. The resulting data series is well fitted by a second-order polynomial function to model the moderator aging, which is the most significant source of positron intensity variation as a function of time. For each moderator regeneration two polynomial fits are performed: the first fitting only the points acquired with the laser on, the second only the points acquired with the laser off. The average of the two resulting fitted polynomials is a model of the evolution of the shot intensity in time and provides the necessary normalization factors. To obtain the $\Area_i^\on$ and $\Area_i^\off$ parameters we divided the measured areas by the value of the average polynomial evaluated at the time each shot was acquired.

\begin{figure}[htp]
\centering
\includegraphics[width=\linewidth]{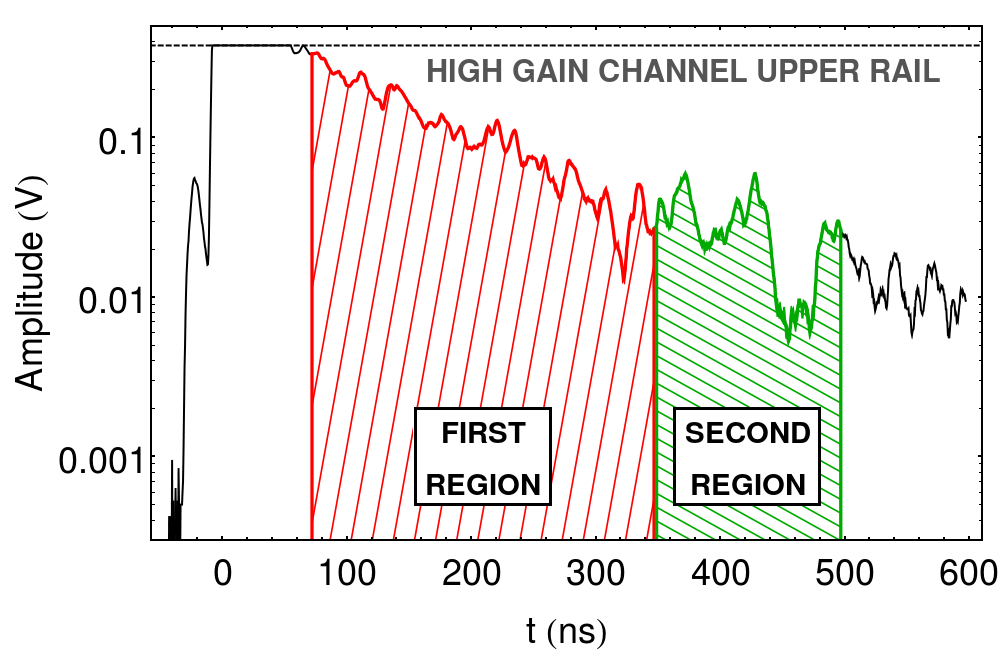}
\caption{SSPALS spectrum of a single shot acquisition with the UV laser on as seen by the high gain channel (continuous black curve). The region around the origin of the horizontal scale where the oscilloscope signal saturates corresponds to the prompt peak. The areas selected for the analysis are highlighted (see text).}
\label{Areas}
\end{figure}

For our analysis of the SSPALS spectra,  two regions were chosen in which the experimental curve was integrated and the $ S_i $ parameter was computed.
The first area was chosen so that it doesn't intersect the prompt peak (which is also implied by the previous condition of the high-gain channel not saturating) and so that most of the produced positronium would still be freely expanding in the chamber without hitting the walls. The range from \SI{70}{\nano\second} to \SI{350}{\nano\second} from the peak satisfies these conditions. Indeed, according to measurements performed on Rydberg Ps in the same chamber and after positron implantation with the same energy and in the same target, the interaction with the walls begins after around \SI{350}{\nano\second} from the prompt peak \cite{NEqual3}. The second region was chosen to lie contiguous to the first one ranging from \SI{350}{\nano\second} to \SI{500}{\nano\second}, where the signal approaches the noise level (see Fig.~\ref{Areas}).

The resulting $S$ parameter obtained considering all of the acquired spectra is computed as

\begin{equation}%
	S ~=~ \frac{\Area_\off - \Area_\on}{\Area_\off} \, ,
	\label{SPDefinition}
\end{equation}

\noindent
where $\Area_\on$ and $\Area_\off$ are the average of the values of $\Area_i^\on$ and $\Area_i^\off$ respectively.

The uncertainty $\Delta S$ of the $S$ parameter, calculated with the described detrending technique, can be derived by using error propagation theory (see Eq.~\ref{ErrorPropagation}) starting from $\Area_\on$, $\Area_\off$ their respective uncertainties $\sigma_\on$ and $\sigma_\off$ (see Eq.~\ref{SigmaDefinition} in Appendix).


\begin{equation}
	\Delta S ~=~ \sqrt{~\frac{\sigma_\on^2}{\Area_\off^2} + \frac{\Area_\on^2 \cdot \sigma_\off^2}{\Area_\off^4}~} \, .
	\label{ErrorPropagation}
\end{equation}

\noindent
Notice that in the whole procedure we avoided any background subtraction.
Indeed we expect the background to be due to the annihilations of Ps inside the nanochannels, due to reemitted positrons and due to the response of the detector to the positron burst, and therefore to be proportional to the shot intensity. As detailed in the Appendix it is counterproductive to attempt a background subtraction under these premises.

\begin{figure*}[htbp]
\centering
\includegraphics[width=0.49 \textwidth]{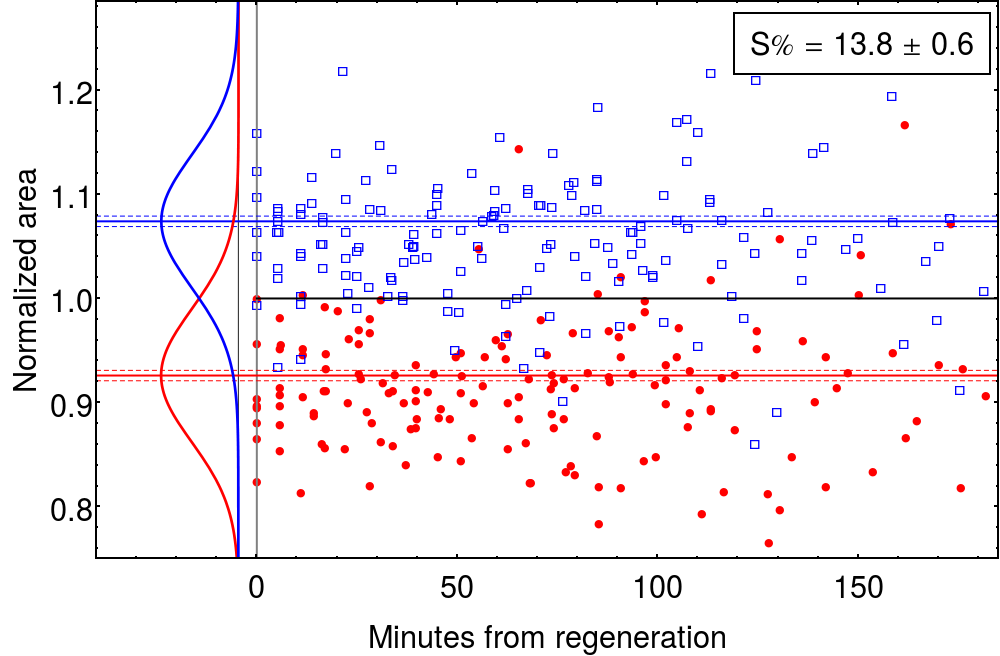}
\includegraphics[width=0.49 \textwidth]{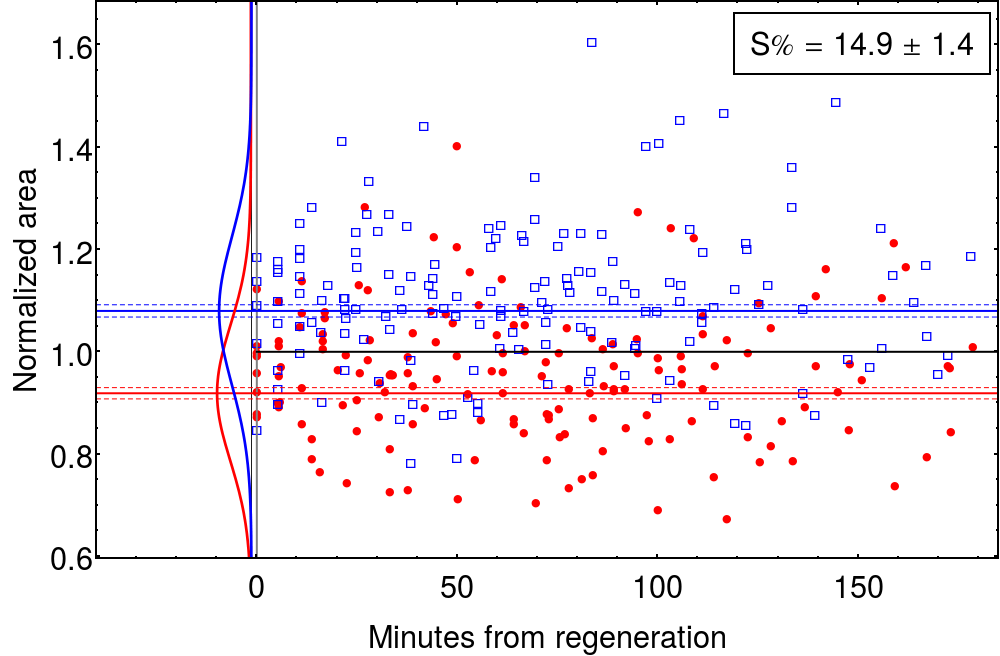}
\caption{Detailed analysis for the run in which both the UV and the IR lasers were both used. Each sample represent the normalized area (\emph{i.e.} $\Area^\on_i$ or $\Area^\off_i$) for a single run. Full circles represent the laser-on runs, empty squares represent laser-off shots. The sideways Gaussian curves represent an estimate of the distribution of the $\Area^{\on/\off}_i$ as described in the text. In the insets the computed values of the $S$ parameter are given.}
\label{UVIRFinal}
\end{figure*}
\begin{figure*}[htbp]
\centering
\includegraphics[width=0.49 \textwidth]{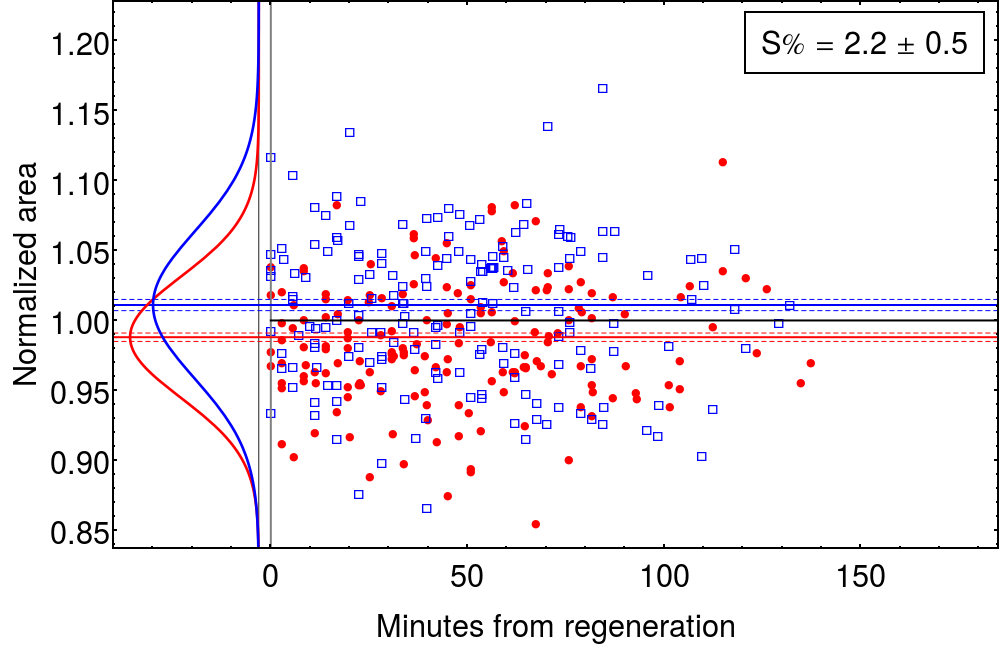}
\includegraphics[width=0.49 \textwidth]{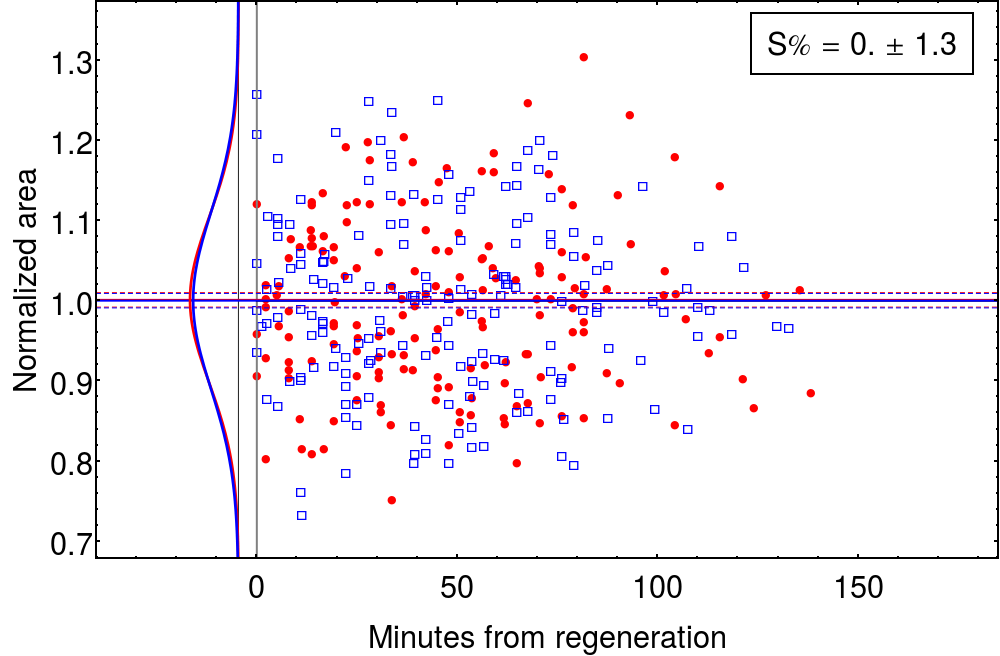}
\caption{Detailed analysis for the run in which only the UV laser was used. Each sample represent the normalized area (\emph{i.e.} $\Area^\on_i$ or $\Area^\off_i$) for a single run. Full circles represent runs in which the laser has been shot, empty squares represent laser-off shots. The sideways Gaussian curves represent an estimate of the distribution of the $\Area^{\on/\off}_i$ as described in the text. In the inset the computed value of the $S$ parameter are given.}
\label{UVFinal}
\end{figure*}

The result of the detrending technique applied to the acquired data on the \emph{first} and \emph{second} regions of the spectra are shown in Fig.~\ref{UVIRFinal} and Fig.~\ref{UVFinal}. The former figure contains data taken with both the UV and IR lasers on, while the latter figure is obtained with the UV laser only. The scatter plot points correspond to single shots, and their horizontal coordinate is given by the time elapsed from the moderator regeneration and the moment when the shot was acquired, while their vertical coordinates correspond to the measured $\Area_i^{\on/\off}$ for each shot. Blue points correspond to shots acquired with the laser off, red points to shots acquired with the laser on.

\begin{table*}
	\begin{tabular}{ccc}
		\textbf{First region} & \hspace{25mm} & \textbf{Second region} \\
		\hline
		\result{0.926}{0.005}{1.074}{0.005}{13.8}{0.6} &
			\textbf{UV + IR} &
			\result{0.919}{0.011}{1.081}{0.012}{14.9}{1.4} \\	 
		\result{0.989}{0.003}{1.011}{0.004}{2.2}{0.5} &
			\textbf{UV} &
			\result{1.001}{0.009}{0.999}{0.009}{0.0}{1.3}
	\end{tabular}
	\caption{Summary of the results of the detrending analysis for the selected areas.}
	\label{ResultTable}
\end{table*}

The two functions plotted sideways in the leftmost part of each graph are Gaussian curves centered on the averages of the $\Area_i^\on$ and $\Area_i^\off$ sets of points, and having the variances of the respective sample distributions. Hence they represent an estimate of the distribution of $\Area_{\on/\off}$ under the assumption that they are normally distributed. The thick horizontal lines also mark the average values $\Area_\on$ and $\Area_\off$, and the dashed envelopes around them range as $\pm \sigma_\on$ and $\pm \sigma_\off$ respectively.

The results of the data analysis are reported in Table~\ref{ResultTable}. They show an $S$ parameter in the \emph{first} region when both lasers are shot (see Fig.~\ref{UVIRFinal}) that is consistent within the experimental uncertainty with the observed $S$ in the \emph{second} region. As mentioned previously, when both lasers are fired a fixed fraction of the produced Ps is removed immediately after the prompt peak and the remaining population retains the original lifetime. The expected behavior, \emph{i.e.} the ratio between the two curves being constant in the SSPALS time frame and not affected by the interaction with the chamber walls, is confirmed.

The UV laser data show a $ 2.2\% $ reduction of the annihilation rate in the \emph{first} region (see Fig.~\ref{UVFinal}) which can be consistently explained with the early stages of the excitation/de-excitation processes, mainly the magnetic quenching. If this were the only phenomenon affecting the Ps lifetime the same value of the $S$ parameter should be observed also in the \emph{second} region, since these processes (as photoionization in the UV+IR case) remove a fraction of the Ps immediately after the prompt peak but do not affect the lifetime of the remaining fraction.
On the contrary, the experimental data show an $S$ value in the \emph{second} region which, given the current precision, is incompatible with that measured in the \emph{first} region with a likelihood ratio \cite{Likelihood} of $90\%$. This variation is instead compatible with the presence of a long-lived fraction of Ps which contributes to the annihilation signal.
This $S$ difference is attributable to the presence of the (partially mixed) metastable \tS{} fraction in the Ps beam. The same conclusion holds by changing the integration intervals by some tens of nanoseconds.

\section{Modelling} \label{SectionIV}

To quantitatively support the interpretation of these results, we formulate here a simplified rate equation model which describes the long time evolution of the (number) populations of Ps atoms in the relevant energy levels, to elucidate the complex dynamics of the optical and annihilation decays after laser excitation.

As discussed in the Introduction, the presence of the fields in the experimental chamber has noteworthy consequences on the optical transition patterns and on the optical decay and annihilation lifetimes of excited Ps atoms. An accurate study of a Ps atom in these conditions can be done with the help of a simulation code which performs, for each $n$ manifold, the diagonalization of the full interaction Hamiltonian in arbitrary electric and magnetic fields \cite{Villa,Caravita}, using the same numerical methodology introduced by \cite{ReducedLifetime}. This code calculates the modified Ps energy levels, the generalized Einstein coefficients for the optical transitions, the sublevel lifetimes and the radiative decay branching ratios.

\begin{figure}[h!tp]
\centering
\includegraphics[width=\linewidth]{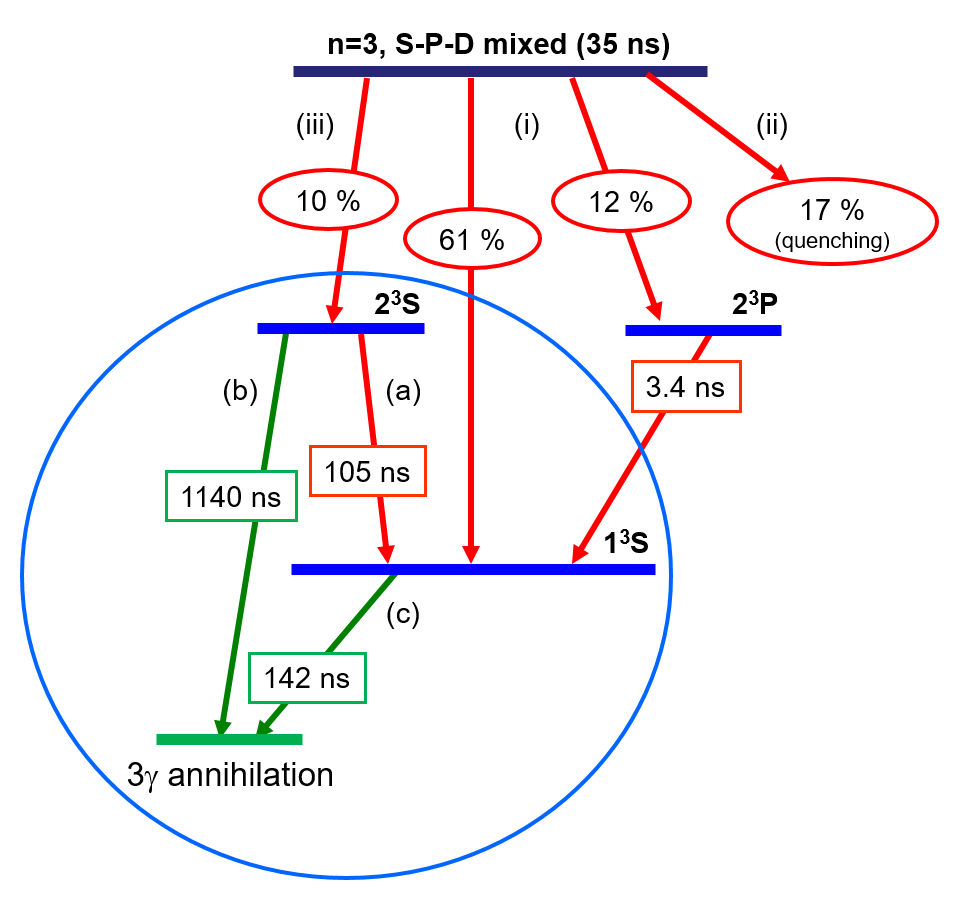}
\caption{Sketch of the relevant energy level structure of a Ps atom after $n = 3$ laser excitation, as discussed in the text. The spontaneous optical decays (red continuous arrows) and annihilation (green dashed arrows) lifetimes are indicated. The small ellipses indicate the calculated branching ratios for the decay of the mixed $n = 3$ sublevels. The large circle encloses the long--lived energy levels and the annihilation patterns considered for the reduced rate equation model.}
\label{scheme}
\end{figure}

Generally speaking, the relevant effects of the fields in our experimental conditions can be summarized as follows (with reference to the energy level diagram shown in Fig.~\ref{scheme}). The presence of the  \SI{25}{\milli\tesla} magnetic field induces some singlet--triplet mixing between states with identical magnetic quantum number $m$, hence some $n =2$ and $n =3$ triplet substates can optically decay towards the singlet ground state \ooS{}, subsequently annihilating with the short lifetime of \SI{125}{ps}. This is what is known as a \emph{magnetic quenching} \cite{PositroniumReview}. On the other hand, the presence of the (average) \SI{300}{\volt \per \centi\meter} electric field induces a Stark effect on the excited Ps atoms, with a mixing between substates belonging to different orbital quantum numbers.

This Stark mixing is stronger for the $n =3$ manifold, where for most substates the S, P or D character of the wavefunctions is completely lost and no known quantum numbers are adequate for their spectroscopic description. The transition probabilities from the $n =3$ sublevels towards those of the $n =2$ and $n =1$ manifolds can be calculated as suitable linear combinations of the transition probabilities between unperturbed sublevels which obey electric dipole selection rules.
Conversely, in the case of the $n =2$ manifold we are in presence of a small to moderate mixing, and the substate wavefunctions mainly retain their unperturbed S or P character, with only a small contribution coming from the wavefunction of the other orbital quantum number. Then, as already implicitly done in the whole paper, we can reasonably designate them again with the unperturbed spectroscopic symbols.

It is important to note, as mentioned in the Introduction, that the small P contribution contained in partially mixed \tS{} states determines a spontaneous decay towards the ground state \otS{}, with emission of one optical photon, otherwise forbidden by electric dipole selection rules. In order to investigate the production of long--lived Ps in such a (partially mixed) \tS{} state, a particular attention was devoted to calculate the \textit{effective optical lifetime} of the whole Ps cloud in our experimental conditions, \emph{i.e.} its characteristic exponential decay time associated with the average spontaneous radiative decay rate \tS{}--\otS{} in the presence of the electric field of our setup. If the electric field was uniform, the \tS{}--\otS{} spontaneous radiative decay rate $ r_{opt} $ would be constant. If, however, the field is not uniform (as in our case), the atom's survival probability can be still approximated (given that the non-uniformities are not too severe) with a negative exponential law where $ r_{opt} $ is the average spontaneous radiative decay rate over the atom's flight trajectory. The average spontaneous radiative decay rate of the whole Ps cloud $ \langle r_{opt} \rangle $ is thus obtained by averaging over all possible trajectories within the cloud, and the effective optical lifetime is given by $ \tau_{opt} = 1 / \langle r_{opt} \rangle $.

\begin{figure*}[htp]
	\centering
	\includegraphics[width=0.48 \textwidth]{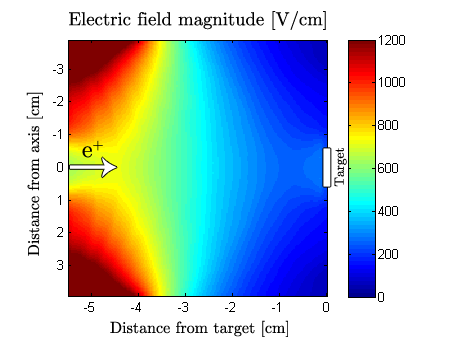}
	\includegraphics[width=0.48 \textwidth]{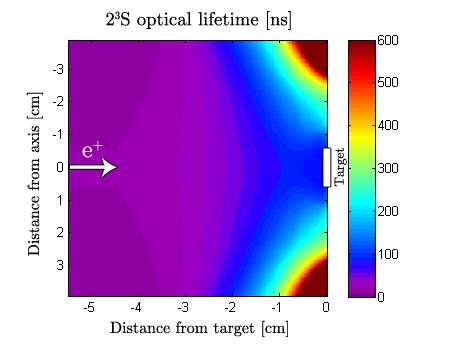}
	\caption{Electric field magnitude in the real experimental geometry (left panel) and optical lifetime of $ 2^3S $ state (right panel, values above \SI{600}{\nano\second} have been clamped for better visual readibility). The target position and the direction of the positron beam are reported. For the detailed geometry of the chamber see \cite{NEqual3,PositronBunching}.}
	\label{MetaLifetime}
\end{figure*}

The calculation of $ \langle r_{opt} \rangle $ in our geometry was performed as follows. A detailed electric field map of our experimental chamber \cite{NEqual3,PositronBunching} was calculated using SIMION \cite{Simion} in the plane orthogonal to the laser propagation axis (see Fig.~\ref{MetaLifetime}, left panel). Using the values of $ r_{opt} $ calculated with the simulation code \cite{Villa,Caravita} as a function of the electric field, a spatial map of the optical lifetime was obtained from the electric field map (see Fig.~\ref{MetaLifetime}, right panel). Finally, the average spontaneous radiative decay rate of the whole \tS{} Ps cloud was calculated by means of a 2D Monte Carlo in the plane orthogonal to the laser propagation axis. Ps was assumed to be emitted isotropically from the target with an axial velocity of $ 1.0 \times 10^5 \, \si{\meter\per\second} $, in agreement with the Doppler velocimetry survey reported in \cite{NEqual3} (conducted in similar experimental conditions with the same target). The trajectory of each atom was assumed to be a straight line with angle $ \theta $ of orientation with respect to the target's normal. A first averaging of the spontaneous radiative decay rate was performed along each trajectory to work out $ r_{opt} $ for each value of $ \theta $. Finally, an averaging over $ \theta $ between $ -\pi/2 $ and $ \pi/2 $ (assuming uniform emission distribution from the nanochannel converter) gave $ \langle r_{opt} \rangle $. This choice is in line with the outcome of \cite{NEqual3} using the same target, where a broad distribution of transverse velocities, consistent with an isotropic Ps emission, was observed.

The effective optical lifetime of the partially mixed \tS{} levels was estimated to be $ \tau_{opt} = \SI{105}{\nano\second} $, essentially due to the increased probability of optical de-excitation as the atoms approach the last electrostatic lens of the bunching system.
Other results derived from our calculations, relevant for the construction of a reliable reduced rate equation model (with reference to Fig.~\ref{scheme}) are: (1) the optical lifetime of the $n =3$ manifold of mixed sublevels amounts to \SI{35}{\nano\second} on average; (2) the optical lifetime of the (partially mixed) \tP{} levels are of the same order of the corresponding unperturbed levels, resulting in \SI{3.4}{\nano\second}. Moreover, we observe that the averaged branching ratio of the metastable \tS{} production from the $n =3$ mixed sublevels can be estimated around to 10\%, while for the \tP{} state production it is 12\%. The remaining 78\% can be attributed to the spontaneous optical decay towards the ground state \otS{} ($\sim$61\%) and to population losses towards rapid annihilating singlet states, mainly due to magnetic quenching ($\sim 17 \%$, see also \cite{NEqual3}).

From these observations it is clear that the long--lived fraction of the excited Ps states determining the long--time behavior of the SSPALS spectra in our experiment is essentially composed of \tS{} states. This suggests to approximately separate the complex population dynamics of excitation plus optical and annihilation decays into two successive parts, as schematically pictured in Fig.~\ref{scheme}.

The first part of the population dynamics starts with the arrival of the laser pulse and lasts a few tens of nanoseconds. The $n =3$ manifold is first populated with efficiency $ \eta_3 \sim 14\% $, according to a weighted average of the photo--ionization experiment data listed in Tab.~1 (and assuming $ \sim 100 \% $ ionization efficiency \cite{NEqual3}). Subsequently, the three processes enumerated in the Introduction take place: (i) rapid optical decay towards the triplet \otS{} ground; (ii) rapid annihilation decay due to magnetic quenching with efficiency $ \eta_q $, representing a net population loss; (iii) spontaneous decay towards the long--lived \tS{} states with relative efficiency $ \eta_m$. These two last quantities will be considered as fitting parameters in the rate equation model.

The second part of the population dynamics starts some tens of nanoseconds after the laser pulse and is dominated by: (a) optical decays of the populated \tS{} states towards the triplet ground state with lifetime $\tau_{opt} = $ \SI{105}{\nano\second}; (b) annihilation decays of \tS{} states into three $\gamma $ photons with lifetime $\tau_2 = $ \SI{1.14}{\micro\second}; (c) annihilation decays of \otS{} with lifetime $\tau_1 =$ \SI{142}{\nano\second}.

A simplified rate equation model of the populations' evolution in the time interval relevant to the experimental data analysis can thus be formulated by disregarding the complex sublevel dynamics right after the laser pulse, and focusing only on the longer time scale level dynamics. Naming $ N_1(t)$ and $ N_2(t)$ the number populations of Ps atoms in \otS{} and \tS{} states, respectively, and introducing $N_0(t)$ as the number population of annihilated Ps atoms, the rate equations which describe the free decay dynamics are:


\begin{equation}
\begin{aligned}
\frac{d N_2}{d t} \, &=  -\frac{N_2(t)}{\tau_2} \,  -\frac{N_2(t)}{\tau_{opt}} \,\\
\frac{d N_1}{d t} \, &= -\frac{N_1(t)}{\tau_1} \, \,+\, \frac{N_2(t)}{\tau_{opt}} \,  \\
\frac{d N_0}{d t} \, &= \frac{N_1(t)}{\tau_1} \, \,+\, \frac{N_2(t)}{\tau_2} \,\,\,\, .
\end{aligned}
\label{EqModel}
\end{equation}

This set of differential equations can be integrated in a straightforward manner by setting proper initial conditions. In the case with laser on, following the above discussion, these are $N_2^\on (t_0) = \eta_3 \times \eta_m $ and $N_1^\on (t_0) = 1 - \eta_3 \times \eta_m - \eta_3 \times \eta_q $. In the case of laser off, where all atoms are assumed to be in the \otS{} states, they are simply $N_1^\off (t_0) = 1$ and $N_2^\off (t_0) = 0$. The initial time of the integration $t_0$ was chosen at $ 16+35 \si{\nano\second} $ after the prompt peak to let the atoms on $ n = 3 $ decay to $ \sim 1 / e $, \emph{i.e.} when the processes in the initial excitation dynamics can start being neglected (as discussed above). Note that in both cases we can arbitrarily set $N_0(t_0) = 0$ because the SSPALS spectra are determined by the derivative $d N_0 / d t$.

Estimated values of the $ S $ parameter were obtained by solving numerically the Eqs.~ \ref{EqModel} and integrating the two sets of solutions with laser on and laser off over the same time intervals used for the experimental data analysis (\emph{i.e.} the first and second region, see Fig.~\ref{Areas}). To obtain the \tS{} production efficiency, the $S$ estimates was fitted to the corresponding experimental results obtained with the UV laser only, by using $ \eta_3 $ and $ \eta_m $ as free parameters, and setting the excitation efficiency to the measured value of $ \eta_3 = 14\% $, as discussed before. The outcome of the fitting procedure was verified by varying the integration starting time in the selected range. The results were found consistent within 5\% and well within their statistical uncertainties.

\begin{table}[h]
\centering
\begin{tabular*}{0.9\columnwidth}{@{\extracolsep{\fill}}lll}
\multicolumn{3}{c}{\textbf{Relative  efficiences}} \\
\hline
$ \eta_m $ & (\tS{} branching eff.) &  $ \  \ ( 14.8 \pm 9.4)\% $ \\
$ \eta_q $ & (quenching eff.) & $ \  \ (15.0 \pm 3.4)\% $ \\
$ \eta_3 \times \eta_m $  & (overall \tS{} prod. eff.) &  $ \  \ (2.1 \pm 1.3) $\%

\end{tabular*}
\caption{Results obtained by fitting the rate equation model (Eqs. \ref{EqModel}) to the experimental data reported in Tab. \ref{ResultTable}. The best--fit parameter values for the 
$ \mathrm{3 \rightarrow 2} $ metastable production efficiency $ \eta_m $ and the quenching probability $ \eta_q $ are those obtained with $t_0 = 16+35 \si{\nano\second} $. The overall metastable production efficiency $ \eta_3 \times \eta_m $ is also reported.}
\label{FitResultTable}
\end{table}


The values of $ \eta_m $ and $ \eta_q $ obtained from the best fit, are reported in Table \ref{FitResultTable}. The estimated \tS{} production efficiency $ \eta_m \simeq (14.8 \pm 9.4)$\% is found statistically compatible with the expected $ 10\% $.
The overall efficiency in exciting triplet ground-state Ps atoms to \tS{} in the present setup is finally obtained by multiplying the two efficiencies, $ \eta_3 \,\times\, \eta_m = (2.1 \,\pm\, 1.3)\% $.  The number of Ps atoms in the \tS{} level produced per $ e^+ $ bunch is obtained from the estimated intensity of our $ e^+ $ source, which delivers on the target $ \sim 1.3 \cdot 10^7 $ $ e^+ $ every \SI{180}{s} (see \cite{PositronBunching}), multiplied by the Ps conversion efficiency of our target ($ \sim 35\% $ see \cite{PositronBunching,NCP}). The estimated production rate of Ps atoms in the \tS{} level is thus $ \sim 1.0 \cdot 10^5 $ every \SI{180}{s}.


\section{Conclusions}

We have studied the excitation of positronium to the long-lived \tS{}
state by spontaneous radiative decay from the \ttP{} level manifold,
excited by a UV laser pulse. The experiment was performed
in a dedicated chamber and in the presence of a guiding $\SI{25}{\milli\tesla}$ magnetic
field and an average $\SI{300}{\volt\per\centi\meter}$ electric field. The presence of
the fields caused mixing between the sublevels of each $n$--manifold,
with the consequence of inducing an optical decay of the populated
\tS{} states (otherwise stable against one--photon radiative decay), finally
shortening the lifetime of these states from $\SI{1.14}{\micro\second}$ to $\SI{105}{\nano\second}$.
Ps atoms in these excited states, anyway, constituted a longer--lived component 
that was observed by means of single--shot positronium annihilation lifetime 
spectra (SSPALS). The evidence
of the successful metastable state production was obtained with a novel
analysis technique of SSPALS data, able to identify very small
deviations from the reference spectra obtained without an
exciting laser pulse.
The experimental results were fitted with a rate equation model which
describes the long--time evolution of the populations of the Ps relevant states after
the $n = 3$ excitation event. Annihilation and
decay rates were obtained using an exact calculation code of Ps
energy levels and optical transitions in arbitrary electric and magnetic fields.

The observed \tS{} state production efficiency relative to the amount of produced Ps was evaluated to be $ \eta_3 \times \eta_m = (2.1 \pm 1.3)$\%. This production efficiency is about 1/3 of the $ 6.2 \% $ recently obtained by Stark mixing between S and P sublevels during \otS{} -- \tP{} laser excitation \cite{AlonsoHoganCassidy}. However, for the Stark mixing method, the rate at which the electric field can be switched off (necessary to avoid the rapid radiative decay of the P component in the mixed state) limits the amount of long--lived \tS{} Ps atoms that survives electric field switching. Conversely, the method demonstrated here is ideally free from this drawback, as it can be realized in absence of any electrical field which induces sublevel mixing and subsequent radiative decay losses.

A further advantage of our method is that it could allow, in a field-free environment, reaching high \tS{} production efficiencies. This can be realized, for example, by increasing the length of the \otS{}--\ttP{} laser pulse to $ > \SI{10}{\nano\second} $ and selecting a bandwidth to optimally cover the transition Doppler profile. After some iterations of laser excitation and spontaneous radiative decay, indeed, a very high fraction of the initial Ps atoms would be pumped to the \tS{} state.
Another possibility could be to add an extra IR laser at \SI{1312.2}{\nano\meter} to directly pump the \ttP{} -- \tS{} transition. Ideally, this would make \otS{} -- \ttP{} -- \tS{} an equally--populated three--level system (assuming no extra losses and saturation of the laser pulses) that could lead to up to $ 10 \% $ excitation efficiency with the present UV bandwidth, corresponding to a fourfold gain with respect to the current setup and start being competitive with direct two-photon excitation (17.6\%, see \cite{Haas}), yet not requiring intense narrow-band lasers.

Thus, the production of Ps atoms in the \tS{} state by spontaneous
radiative decay from the \ttP{} in the absence of external fields, seems to be a promising
alternative for obtaining this metastable Ps state, potentially leading
to higher production efficiency.
Measurements of the \ttP{} -- \tS{} decay in a electric field free environment
are planned in order to verify the suitability of this method.


\appendix{}

\section{The detrending technique}
\label{DetrendingApp}

\begin{figure*}[htp]
\centering
\includegraphics[width=0.7 \linewidth]{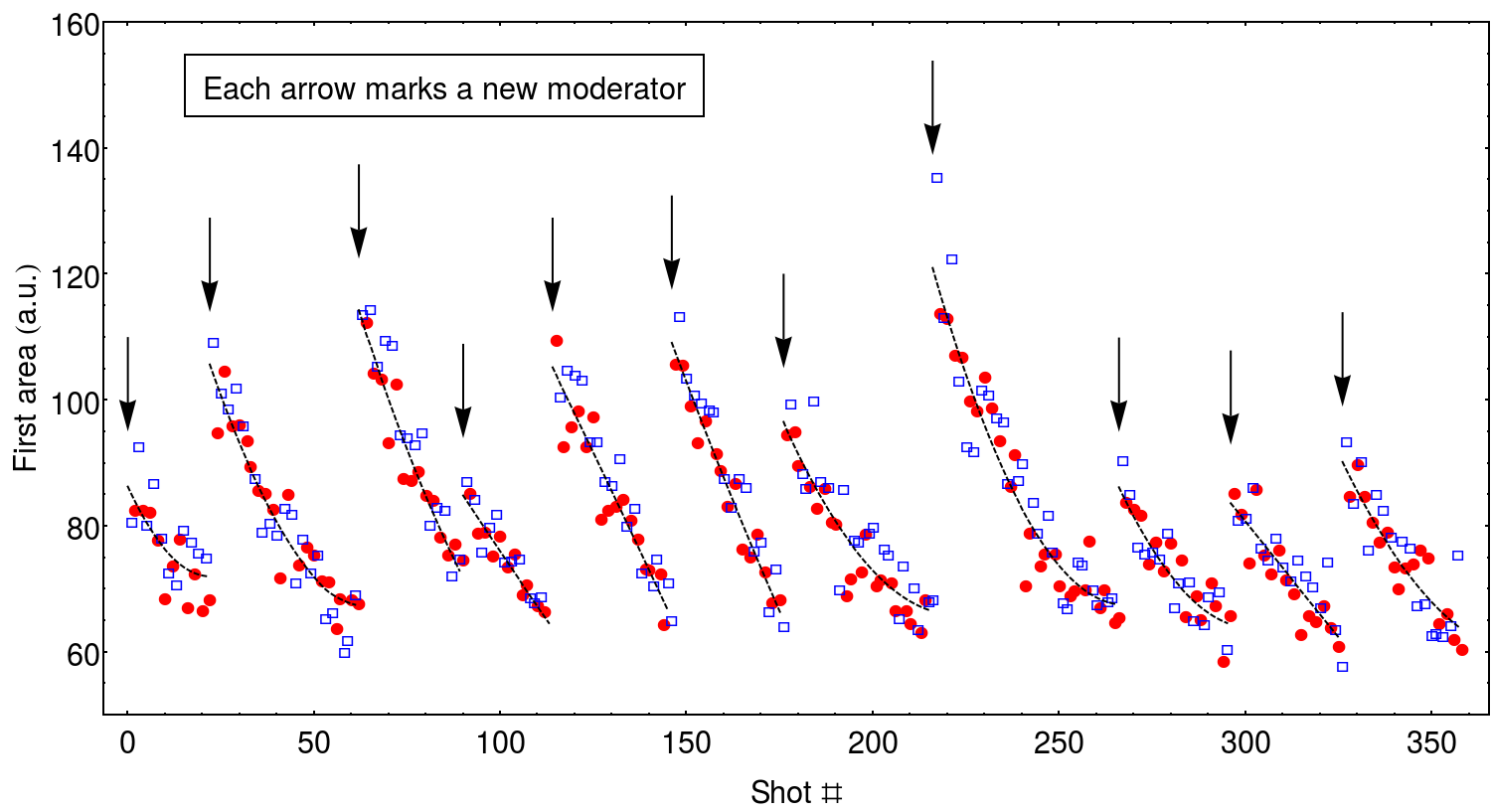}
\caption{Value of the free SSPALS spectrum area in the UV laser shots, their corresponding laser-off counterparts and the best-estimation of the normalization factor $ f(t_i) $ (see Eq.~\ref{FDefinition}, where $ t_i $ is the time at which each shot was acquired) plotted against the progressive number of each shot. Full circles represent laser-off measurements, open squares represent laser-on measurements, the dashed lines are the second-order polynomials resulting from the fit of each group of shots related to a different moderator (see text). No normalization was performed to compensate variations of the beam intensity. The decrease in the shot intensity due to the moderator aging is clearly visible.}
\label{Progression}
\end{figure*}

Let us consider a single group of shots, \emph{i.e.} a single series of alternating laser-on and laser-off measurements taken after a single moderator regeneration. The consequence of the decrease in positron beam intensity following each moderator regeneration is evident in each group of shots shown in Fig.~\ref{Progression}. We designate with $\widetilde I(t)$ the beam intensity at the time $t$, where $t$ is the time elapsed since the moderator regeneration; hence $\widetilde I(t)$ is the number of positrons that would hit the target if a shot was fired at $t$. Here and in the following we use the tilde notation to indicate exact values as opposed to their approximation obtained through experimental measurements or by means of empiric formulae.

Then, let us consider a parameter $ \widetilde P $, linear function of the ideal SSPALS profile (and therefore, implicitly a function of time) such as the intensity of the SSPALS at a specific point, or the area under the SSPALS profile in a specific region. Since, reasonably, we expect ideal SSPALS profiles to be proportional to the beam intensity (but dependent on the laser on/off condition), $\widetilde P$ will be a linear function of $\widetilde I$ with two separate coefficients for the laser-on and laser-off condition. Calling these coefficients $\widetilde k_\on$ and $\widetilde k_\off$ we can write

\begin{equation}
	\widetilde P_{\on/\off}(t) = \widetilde k_{\on/\off} \widetilde I(t)
	\label{KDefinition}
\end{equation}
and compute the average $\widetilde P$
\begin{equation}
	\widetilde f(t) = \frac{\widetilde k_\on + \widetilde k_\off}{2} \, \widetilde I(t)
	\label{FDefinition}
\end{equation}

\noindent
which, among all of the quantities proportional to $\widetilde I(t)$, can be approximated with high precision from the experimental data. To calculate its best possible approximation ${f}(t)$ we start by computing the experimental value of the $\widetilde P$ parameter over all measured SSPALS spectra in the shot group. Let $t^\on_i$ be the time at which the $i^\text{th}$ shot with the laser on was acquired and $P^\son_i$ the value of the parameter $\widetilde P$ computed over the SSPALS spectrum of such shot. Let $t^\off_i$ and $P^\off_i$ be similarly defined for the shots acquired with the laser off.

We now assume (as is reasonable in our specific case) that $\widetilde I(t)$ can be well approximated with a suitable empiric formula $I(t)$. This implies that also any linearly dependent quantity of the intensity $\widetilde I(t)$ can be approximated with such model, at most with the introduction of a multiplicative factor. We approximate $\widetilde f(t)$ by fitting separately the $\{ t^\on_i, P^\son_i\}$ and the $\{ t^\off_i, P^\off_i\}$ datasets with the $I$ model, then compute the arithmetic mean of the two fitted functions, thus obtaining $f(t)$.

Since our final goal is to normalize the parameter $\widetilde P(t)$ to the beam intensity (or to something proportional to it) we now define the ratio between the value of the parameter $\widetilde P(t)$ and the function $\widetilde  f(t)$

\begin{equation}
	\widetilde \mu_\on ~=~ \frac{\widetilde P_{\on}(t)}{\widetilde f(t)}
	\label{MuTildeDefinition}
\end{equation}

\noindent
and similarly $\widetilde \mu_\off$. By applying the definition of $\widetilde P(t)$ (Eq.~\ref{KDefinition}) and $\widetilde f(t)$ (Eq.~\ref{FDefinition}) it follows that $\widetilde \mu_\on$ is independent of $t$:

\begin{equation}
	\widetilde \mu_\on ~=~ \frac{\widetilde P_\on(t)}{\widetilde f(t)} ~=~ \frac{2 \widetilde k_\on}{\widetilde k_\on + \widetilde k_\off} \, .
	\label{MuIndependent}
\end{equation}

\noindent
Moreover, it follows directly from the given definitions (Eq.~\ref{KDefinition}, \ref{FDefinition} and \ref{MuIndependent}) that $P^\son_i / f(t^\on_i)$ approximates $\widetilde \mu_\on$:

\begin{equation}
	\widetilde \mu_\on = \frac{2 \widetilde k_\on}{\widetilde k_\on + \widetilde k_\off} \approx \frac{2}{\widetilde k_\on + \widetilde k_\off} \frac{ P^\son_i }{ I(t^\on_i) } = \frac{P^\son_i }{f(t^\on_i) } \, .
	\label{MuMeaning}
\end{equation}

\noindent
Therefore the arithmetic mean of the $P_i^\son / f(t^\on_i)$ gives the best approximation of $\widetilde \mu_\son$, and we designate it $\mu_\son$:

\begin{equation}
	\mu_\on ~=~ \left< \frac{ {P}^\son_i }{ {f}(t^\on_i) } \right> \, .
	\label{MuDefinition}
\end{equation}

\noindent
The final averaging operation is permitted by the independence of $\widetilde \mu_\on$ from time. The same applies, of course, to $\widetilde \mu_\off$ and its approximation $\mu_\off$.



Assuming that there is no residual trending in the points (\emph{i.e.} the probability distribution of the $P^\son_i / f(t^\on_i)$ is independent of $t$) the uncertainty of the estimation of the parameter $\widetilde \mu_\on$ can be computed as:

\begin{equation}
		\sigma_\on = \sqrt{~ \frac{1}{N_{dof}} \left< \left( \frac{P^\son_i}{f(t^\son_i)} - \mu_\on  \right)^2 \right> ~} \, ,
		\label{SigmaDefinition}
\end{equation}

\noindent
where $N_{dof} = N_{acq} - N_{par} $ is the number of degrees of freedom given $N_{par}$ is the number of free parameters in the intensity evolution model $I(t)$ and $N_{acq}$ is the number of laser-on acquisitions in the family.

If, based on the experimental setup, we can formulate an error model that assigns as uncertainty $\Delta P^{\son/\off}_i$ to each measurement, then the mean in Eq.~\ref{MuDefinition} and Eq.~\ref{SigmaDefinition} should be weighted according to these uncertainties. It is not required of the error model to give a correct estimation of the magnitude of the uncertainties $\Delta P^{\son/\off}_i$ but only of their relative ratio, since only that is relevant when computing the weighted mean.

Caution must be taken when considering the opportunity of attempting a background subtraction on the measured SSPALS spectra before applying the technique.
If the background is, in fact, expected to be proportional to the shot intensity, performing its subtraction should be ideally equivalent to multiplying all areas by a constant factor $\alpha$; in practice it will be subject to the experimental errors deriving from background measurement and the estimation of its correct normalization.

We can see that arithmetically the result expected from the background-subtracted spectra is no different from the one obtained without background subtraction. This means that performing the background subtraction does not temper or eliminate systematic errors. Let us use a hat ($\, \hat{} \,$) to indicate the parameters that we employed early on, but this time computed after subtracting the background:

\begin{equation}
\begin{aligned}
	{\hat{\widetilde{P}}_i^{\son / \off}} &= \alpha \tilde P^{\son / \off}_i \\
	\hat{\widetilde{f}}(t) &= \alpha \tilde f(t) \, .
\end{aligned}
\end{equation}

\noindent
Therefore, from Eq.~\ref{MuTildeDefinition}:

\begin{equation}
\begin{aligned}
	\Rightarrow ~~~~ \hat{\widetilde{\mu}}_{\son / \off} &= \frac{\hat{\widetilde{P}}_{\son / \off}}{\hat{\widetilde{f}}_{\son / \off}(t)} \\
	&= \frac{\tilde P_{\son / \off}}{\tilde f_{\son / \off}(t)} ~=~ \tilde \mu_{\son / \off} \, .
\end{aligned}
\end{equation}

\noindent
The ratio of $\hat{P}_\son$ and $\hat{f}_\son(t)$, background-subtracted measured values, is an estimator of the same value as the ratio of ${P}_\son$ and ${f}_\son(t)$, non background-subtracted measured values, albeit subject to the additional uncertainties deriving from the process of background subtraction. As a consequence it will always be better in terms of the final uncertainty to employ the ratio of ${P}_{\son / \off}$ and ${f}_{\son / \off}(t)$ to compute $\mu_{\son / \off}$.

When we applied the detrending analysis on the data discussed in this article, we employed a second-order polynomial function as the model used to fit the intensity evolution and the area under the curve in a chosen window  as the parameter $P$. Hence the quantities of $\Area_{\on/\off}$ used to define the $S$ parameter become interchangeable with the quantities $\mu_{\on / \off}$. Therefore:

\begin{equation}
\begin{aligned}
	\frac{\mu_\off - \mu_\on}{\mu_\off} &= \frac{{P_\off(t)}/{f(t)} - {P_\on(t)}/{f(t)}}{{P_\off(t)}/{f(t)}} \\
	&= \frac{\Area_\off - \Area_\on}{\Area_\off} ~\equiv~ S \, .
\end{aligned}
\label{SComputation}
\end{equation}




\noindent
The detrending technique as described above applies to shots coming from a single group. One could be tempted to detrend data coming from different groups with a common $f(t)$. Instead, since the exact decay profile will differ from shot to shot each group of shots needs to be detrended separately (as shown in Fig.~\ref{Progression}). This is due to two factors: first that each regeneration yields a slightly differently performing Ne moderator; second that its aging, in particular in the first few minutes after regeneration, is quite steep. Thus there is a strong dependence of the initial efficiency on the time elapsed between the regeneration and the first shot. After the $f(t)$ of each group has been computed, $\mu_\on$ can be computed as the average of all the $P_i^\son / f(t_i^\son)$ coming from the different groups. This is due to $\widetilde \mu_\on$ being independent of $t$, and therefore of $\widetilde I(t)$ (see Eq.~\ref{MuIndependent}).

\section*{Acknowledgments}

This work was supported by Istituto Nazionale di Fisica Nucleare; the CERN Fellowship programme and the CERN Doctoral student programme; the Swiss National Science Foundation Ambizione Grant (No. 154833); a Deutsche Forschungsgemeinschaft research grant; an excellence initiative of Heidelberg University; Marie Sklodowska-Curie Innovative Training Network Fellowship of the European Commission's Horizon 2020 Programme (No. 721559 AVA); European Research Council under the European Unions Seventh Framework Program FP7/2007-2013 (Grants Nos. 291242 and 277762); Austrian Ministry for Science, Research, and Economy; Research Council of Norway; Bergen Research Foundation; John Templeton Foundation; Ministry of Education and Science of the Russian Federation and Russian Academy of Sciences and the European Social Fund within the framework of realizing the project, in support of intersectoral mobility and quality enhancement of research teams at Czech Technical University in Prague (Grant No. CZ.1.07/2.3.00/30.0034).

\end{document}